\documentclass[12pt]{article}
\begin{document}
\pagenumbering{arabic}
\title{Gravitational energy-momentum flow in binary systems}

\author{J. W. Maluf$\,^{(1,a)}$, S. C. Ulhoa$\,^{(2,b)}$ 
and J. F. da Rocha-Neto$\,^{(1,c)}$}
\date{}
\maketitle

{\footnotesize
\noindent{\it  (1) Instituto de F\'{\i}sica, Universidade de 
Bras\'{\i}lia, C.P. 04385, 70.919-970 Bras\'{\i}lia DF, Brazil}\par
\bigskip
\noindent{\it (2) Faculdade UnB Gama, Universidade de Bras\'ilia,
72.405-610, Gama DF, Brazil}}
\bigskip

\begin{abstract}
We investigate the gravitational energy-momentum distribution in the space-time
of two black holes in circular orbit, in the context of the teleparallel 
equivalent of general relativity. This field configuration is important because
gravitational waves are expected to be emitted in the final stages of inspiral 
and merger of binary black holes. We address an approximate solution of 
Einstein's field equations that describes two non-spinning black holes that 
circle each other in the $xy$ plane, obtain the total energy of the space-time 
and verify that the gravitational binding energy  is negative. We show
that gravitational radiation is emitted as long as the separation between the
holes decreases in time. If the black holes are spinning and circle each other,
it has been found in the literature
that, during the pre-merger inspiral, they bob up and down 
sinusoidally. The understanding of this phenomenon requires the understanding 
of the gravitational energy-momentum flow in the space-time of binary black 
holes. For the time dependent metric tensor of a general binary
black hole system, the non-vanishing of the gravitational momentum may explain
the bobbing of spinning black holes.
\end{abstract}
\par

\bigskip
{\footnotesize
\noindent (a) wadih@unb.br, jwmaluf@gmail.com\par
\noindent (b) sc.ulhoa@gmail.com\par
\noindent (c) rocha@fis.unb.br\par}

\bigskip
\section{Introduction}
The physics of black holes coalescence is presently being 
intensively investigated. So far it is not known any exact solution of 
Einstein's field equations that describes the inspiral and merger of black 
holes. The progress in this field is due to post-Newtonian approximation 
methods and to nonlinear numerical simulations of the evolution of 
binary black hole space-times 
\cite{Pretorius,Campanelli,Pretorius2}. It is expected that 
binary black hole mergers will provide important information about the 
strong-field, nonlinear nature of the gravitational field, and will become a 
promising source of gravitational waves. Effects like radiation of mass 
(energy), linear and angular momentum are likely to take place in the final 
stages of black hole mergers. In particular, radiation of linear momentum by a
binary black hole is related to the recoil of the final remnant hole (see refs. 
\cite {Campanelli2,Campanelli3,Gonzalez,Blanchet} and references therein). If 
the remnant black hole acquires linear momentum, the latter should cancel with 
the linear momentum of the field in order to comply with the conservation of 
the total linear momentum of the configuration. 

An intriguing and interesting phenomenon is the orbital motion of two 
identical, spinning black holes in quasi-circular orbit, with oppositely
directed spins restricted to the orbital plane 
\cite{Campanelli3,Thorne,Lovelace}. 
During the pre-merger inspiral the orbital plane of the binary black holes 
(plane $xy$, say) carry out a movement up and down along the $z$ direction, 
i.e., the two black holes bob up and down sinusoidally and synchronously. 
After the merger the remnant black hole acquires a recoil velocity, and in 
realistic situations it may be ejected from the nucleous of the host galaxy
\cite{Campanelli2,Campanelli3,Blanchet}. 

The conservation of the total linear momentum of the space-time implies
that the linear momentum of the field must be the same in magnitude, but 
opposite in 
sign, to the linear momentum of the binary/merged black holes \cite{Thorne}. 
Therefore the description of the linear momentum of the gravitational field 
in the space-time of binary black holes is important for the understanding of
the physics of this configuration. Ideally one would like to know the details
of the momentum flow between the fields and holes. These issues are 
mathematically intricate, but in principle they can be addressed (at least
formally) in the framework of the teleparallel equivalent of general 
relativity (TEGR), provided a realistic post-Newtonian expression for the metric
tensor is known. In the TEGR the expressions for the energy-momentum and
angular momentum of the gravitational field are invariant under transformations
of the coordinates
of the three-dimensional spacelike surface, under time reparametrizations, but
depend on the frame of an observer. But normally the observer is stationary in 
the asymptotically flat space-time. The energy-momentum of any physical system 
in special relativity depends on the frame of an observer, and there is no 
special reason for dropping this feature when considering general relativity. 
The gravitational energy is the zero component of the gravitational 
energy-momentum four-vector, and thus it has standard transformation 
properties.

In this paper, we first address an approximate solution \cite{Alvi,Owen} of 
Einstein's equations that describes two nonspinning black holes in 
circular motion, and evaluate the gravitational energy-momentum of the 
space-time. In this 
approximate model the orbital motion of the holes is restricted to the $xy$ 
plane and the separation $b$ between the holes is considered fixed (in the
context of ref. \cite{Alvi}). For two 
black holes with individual rest masses $m_1$ and $m_2$, we will find that the 
average value (in time) of the total energy $cP^{(0)}$ of the combined system 
in orbital motion is less than $c(P^{(0)}_1+P^{(0)}_2)=(m_1+m_2)c^2$. The 
binding energy $E_b=cP^{(0)}-c(P^{(0)}_1+P^{(0)}_2)$ is negative, its 
expression is very simple and is in agreement with previous analyses. 
Assuming that the separation $b$
between the two black holes decreases in time, i.e., $\dot{b}<0$, the
binding energy yields a positive flux of gravitational radiation. 
For a fixed value of $b$ we arrive at a simple expression for the total flux 
of gravitational radiation. 
We will show that the average value of the radiation over a period is zero. 
The conclusion is that effective gravitational radiation takes place provided
$\dot{b}\ne 0$.

We also consider the general form of the metric tensor that describes the
space-time of two spinning black holes in quasi-circular motion in the $xy$ 
plane \cite{Thorne,Faye}, and evaluate the gravitational energy-momentum of the
space-time.
We obtain formal expressions for the linear momentum of the field along the 
$x$, $y$ and $z$ directions, contained within a large rectangular volume with 
sides $a$. This length is supposed to be much larger than the separation $b$ 
between the two black holes. We find that, in general, the linear momenta 
vary with time. The dependence in time of the 
$z$ component of the linear momentum is likely to be related to the bobbing
of realistic binary black holes, prior to the merger. In our opinion, the 
present approach is better suited for this analysis, rather than the one based 
on pseudotensors \cite{Thorne}. Pseudotensors are quantities that depend on the
choice of the coordinates of the three-dimensional space, and therefore they 
are not well behaved under coordinate transformations. The analysis developed
in ref. \cite{Thorne} makes use of the Landau-Lifshitz pseudotensor 
\cite{Landau}. However, we have never seen a justification as to why one 
pseudotensor is better than another one.

This paper is organized as follows. In section 2, we review the formulation of
the TEGR, and show how the definition of the gravitational energy-momentum 
arises out of the field equations of the theory. We also present the expression
for the fluxes of gravitational radiation and radiation of matter fields. 
In section 3, we describe the approximate solution for two black holes in 
circular orbit, and in section 4 we evaluate the energy and momentum of the 
gravitational field. We find simple expressions for the total gravitational 
energy, for the binding energy and for the total flux of gravitational 
radiation. In section 5, we consider the general form of the metric tensor for
the inspiral of two spinning black holes that circle each other, and obtain the
formal expressions for the gravitational momenta along the three spatial 
directions. Assuming the standard asymptotic behaviour of the metric tensor
components, it will be clear that the momentum components of the gravitational
field are time dependent, a fact that very likely explains the bobbing of the 
black holes. Finally we present our conclusions in section 6.

\bigskip
Notation: space-time indices $\mu, \nu, ...$ and SO(3,1) indices $a, b, ...$
run from 0 to 3. Time and space indices are indicated according to
$\mu=0,i,\;\;a=(0),(i)$. The tetrad field is denoted $e^a\,_\mu$, and the 
torsion tensor reads $T_{a\mu\nu}=\partial_\mu e_{a\nu}-\partial_\nu e_{a\mu}$.
The flat, Minkowski spacetime metric tensor raises and lowers tetrad indices 
and is fixed by $\eta_{ab}=e_{a\mu} e_{b\nu}g^{\mu\nu}= (-1,+1,+1,+1)$. The 
determinant of the tetrad field is represented by $e=\det(e^a\,_\mu)$.
\bigskip

\section{Energy-momentum in the TEGR}
In the teleparallel equivalent of general relativity the gravitational field 
is represented by the tetrad field $e^a\,_\mu$ only, and the Lagrangian density
is written in terms of the torsion tensor 
$T_{a\mu\nu}=\partial_\mu e_{a\nu}-\partial_\nu e_{a\mu}$. This tensor is 
related to the antisymmetric part of the Weitzenb\"ock connection 
$\Gamma^\lambda_{\mu\nu}=e^{a\lambda}\partial_\mu e_{a\nu}$. However, the
dynamics of the gravitational field in the TEGR is essentially the same as in
the usual metric formulation. The physics in both formulations is identically 
the same.

Let us start with the torsion-free, Levi-Civita connection $^0\omega_{\mu ab}$,

\begin{eqnarray}
^0\omega_{\mu ab}&=&-{1\over 2}e^c\,_\mu(
\Omega_{abc}-\Omega_{bac}-\Omega_{cab})\,, \\ \nonumber
\Omega_{abc}&=&e_{a\nu}(e_b\,^\mu\partial_\mu
e_c\,^\nu-e_c\,^\mu\partial_\mu e_b\,^\nu)\,.\label{1}
\end{eqnarray}
The Christoffel symbols ${}^0\Gamma^\lambda_{\mu\nu}$ and the
Levi-Civita connection are identically related by 

$$^0\Gamma^\lambda_{\mu\nu}=e^{a\lambda}\partial_\mu e_{a\nu}+
e^{a\lambda}\,(^0\omega_{\mu ab})e^b\,_\nu\,.$$ 
In view of this expression an identity arises between the Levi-Civita 
connection and the contorsion tensor $K_{\mu ab}$, 

\begin{equation}
^0\omega_{\mu ab}=-K_{\mu ab}\,, \label{2}
\end{equation}
where $K_{\mu ab}=\frac{1}{2}e_{a}\,^{\lambda}e_{b}\,^{\nu}
(T_{\lambda\mu\nu}+T_{\nu\lambda\mu}+T_{\mu\lambda\nu})$, and 
$T_{\lambda\mu\nu}= e^a\,_\lambda T_{a\mu\nu}$.
Making use of eq. (2) it follows that the scalar curvature $R(e)$ may be
identically written as

\begin{equation}
eR(^0\omega) =-e\left({1\over 4}T^{abc}T_{abc}+{1\over
2}T^{abc}T_{bac}-T^aT_a\right) +2\partial_\mu(eT^\mu)\,,\label{3}
\end{equation}
where $e$ is the determinant of the tetrad field. Therefore in the framework
of the TEGR the Lagrangian density for the gravitational and matter fields is
defined by

\begin{eqnarray}
L&=& -k e({1\over 4}T^{abc}T_{abc}+{1\over 2}T^{abc}T_{bac}-
T^aT_a) -{1\over c}L_M \nonumber \\
&\equiv& -ke\Sigma^{abc}T_{abc} -{1\over c}L_M\,, 
\label{4}
\end{eqnarray}
where $k=c^3/16\pi G$, $T_a=T^b\,_{ba}$, 
$T_{abc}=e_b\,^\mu e_c\,^\nu T_{a\mu\nu}$ and

\begin{equation}
\Sigma^{abc}={1\over 4} (T^{abc}+T^{bac}-T^{cab})
+{1\over 2}( \eta^{ac}T^b-\eta^{ab}T^c)\;.
\label{5}
\end{equation}
$L_M$ stands for the Lagrangian density for the matter fields. 
The Lagrangian density $L$ is invariant under the global SO(3,1) group. The 
absence in the Lagrangian density of the divergence term on the right hand
side of eq. (3) prevents the invariance of (4) under arbitrary local SO(3,1)
transformations.

The field equations derived from (4) are equivalent to Einstein's equations. 
They read

\begin{equation}
e_{a\lambda}e_{b\mu}\partial_\nu (e\Sigma^{b\lambda \nu} )-
e (\Sigma^{b\nu}\,_aT_{b\nu\mu}-
{1\over 4}e_{a\mu}T_{bcd}\Sigma^{bcd} )={1\over {4kc}}eT_{a\mu}\,,
\label{6}
\end{equation}
where
$\delta L_M / \delta e^{a\mu}=eT_{a\mu}$. From now on we will make $c=1=G$.

The definition of the gravitational energy-momentum may be established in the
framework of the Lagrangian formulation defined by (4), according to the 
procedure of ref. \cite{Maluf1}. Equation (6) may be rewritten as 

\begin{equation}
\partial_\nu(e\Sigma^{a\lambda\nu})={1\over {4k}}
e\, e^a\,_\mu( t^{\lambda \mu} + T^{\lambda \mu})\;,
\label{7}
\end{equation}
where $T^{\lambda\mu}=e_a\,^{\lambda}T^{a\mu}$ and
$t^{\lambda\mu}$ is defined by

\begin{equation}
t^{\lambda \mu}=k(4\Sigma^{bc\lambda}T_{bc}\,^\mu-
g^{\lambda \mu}\Sigma^{bcd}T_{bcd})\,.
\label{8}
\end{equation}
In view of the antisymmetry property 
$\Sigma^{a\mu\nu}=-\Sigma^{a\nu\mu}$ it follows that

\begin{equation}
\partial_\lambda
\left[e\, e^a\,_\mu( t^{\lambda \mu} + T^{\lambda \mu})\right]=0\,.
\label{9}
\end{equation}
The equation above yields the continuity (or balance) equation,

\begin{equation}
{d\over {dt}} \int_V d^3x\,e\,e^a\,_\mu (t^{0\mu} +T^{0\mu})
=-\oint_S dS_j\,
\left[e\,e^a\,_\mu (t^{j\mu} +T^{j\mu})\right]\,,
\label{10}
\end{equation}
Therefore we identify
$t^{\lambda\mu}$ as the gravitational energy-momentum tensor \cite{Maluf1},

\begin{equation}
P^a=\int_V d^3x\,e\,e^a\,_\mu (t^{0\mu} 
+T^{0\mu})\,,
\label{11}
\end{equation}
as the total energy-momentum contained within a volume $V$ of the 
three-dimensional space,

\begin{equation}
\Phi^a_g=\oint_S dS_j\,
\, (e\,e^a\,_\mu t^{j\mu})\,,
\label{12}
\end{equation}
as the gravitational energy-momentum flux \cite{Maluf2}, and

\begin{equation}
\Phi^a_m=\oint_S dS_j\,
\,( e\,e^a\,_\mu T^{j\mu})\,,
\label{13}
\end{equation}
as the energy-momentum flux of matter. In view of (7) eq. (11) may be written
as 

\begin{equation}
P^a=-\int_V d^3x \partial_j \Pi^{aj}=-\oint_S dS_j\,\Pi^{aj}\,,
\label{14}
\end{equation}
where $\Pi^{aj}=-4ke\,\Sigma^{a0j}$. The expression above is the definition 
for the gravitational energy-momentum presented in ref. \cite{Maluf4}, 
obtained in the framework of the vacuum field equations in Hamiltonian form. 
It is invariant under coordinate transformations of the three-dimensional space
and under time reparametrizations. We note that (9) is a true energy-momentum 
conservation equation.

Finally we remark that in the absence of matter fields the total flux of 
gravitational radiation $\Phi^{(0)}$ is related to the total gravitational 
energy according to $\Phi^{(0)}=-\dot{P}^{(0)}$, in view of eq. (10).

\section{The space-time of the binary black hole}

We consider the approximate metric for the binary black hole as described
in ref. \cite{Alvi}. This solution was later re-analyzed and re-obtained 
in ref. \cite{Owen} by matching two perturbed Schwarzschild metrics to an 
asymptotically post-Newtonian construction for a binary black hole 
space-time. The two black holes have masses $m_1$ and $m_2$, and
circle around each other in the plane $xy$. We restrict 
the analysis to the coordinates in the radiation zone, defined by
$r>> \lambda_{GW}$ \cite{Yunes1,Yunes2}, where $\lambda_{GW}$ is the 
wavelength of the gravitational radiation. In the context of the present 
analysis the radiation zone is established in an equivalent way by 
$r>>m_1$, $r>>m_2$ and $r>> b$ (see below the definition of $b$). 
We define

\begin{equation}
m=m_1+m_2\,, \,\,\,\,\,\delta m=m_1-m_2\,,\,\,\,\,\,\, \mu={{m_1m_2}\over m}\,.
\label{15}
\end{equation}

The circular, Newtonian trajectories of the black holes are

\begin{equation}
{\bf r}_1(t)={m_2 \over m}{\bf b}(t)\,, \,\,\,\,\,\,\,\,
{\bf r}_2(t)=-{m_1 \over m}{\bf b}(t)\,,
\label{16}
\end{equation}
where

\begin{equation}
{\bf b}(t)={\bf r}_1(t)-{\bf r}_2(t)=b(\cos\omega t, \sin\omega t, 0)\,,
\label {17}
\end{equation}
and

$$\omega=\sqrt{ {m\over b^3}}\,,$$
is the orbital angular velocity. The separation $b$ is defined by 
$b=|{\bf r}_1-{\bf r}_2|$. The velocity of the holes are

\begin{equation}
{\bf v}_1={{d {\bf r}_1}\over{dt}}\,,\,\,\,\,
{\bf v}_2={{d {\bf r}_2}\over{dt}}\,,
\label{18}
\end{equation}
from what we define 

\begin{equation}
{\bf v}(t)
={\bf v}_1-{\bf v}_2=\sqrt{m\over b}(-\sin\omega t,\cos\omega t, 0)\,.
\label{19}
\end{equation}

In the radiation zone the metric components depend on ${\bf b}$ and ${\bf v}$,
which in turn depend not exactly on $t$, but on the retarded time $\tau=t-r$ 
\cite{Alvi}.
We further define

\begin{equation}
\tilde{m}=m\biggl(1-{\mu \over {2b}}\biggr)\,,\,\,\,\,\,\,
{\bf n}={{\bf r}\over r}\,,
\label{20}
\end{equation}
where ${\bf r}$ is a point of observation in space.

With the help of the definitions above the metric tensor components for the 
binary black hole space-time in the radiation zone read \cite{Alvi}

\begin{eqnarray}
g_{00}&=&-1+{{2\tilde{m}}\over r}-{{2\tilde{m}^2\over r^2}} \nonumber \\
&{}&+ {\mu \over r}\biggl\{2({\bf n}\cdot{\bf v})^2-
{{2m}\over b^3}({\bf n}\cdot{\bf b})^2+
{6\over r}({\bf n}\cdot{\bf b})({\bf n}\cdot{\bf v})+
{1\over r^2}\lbrack 3({\bf n}\cdot{\bf b})^2-b^2\rbrack \biggr\}\nonumber \\
&{}&+{\mu \over r} {{\delta m}\over m}\biggl\{({\bf n}\cdot{\bf v})\biggl[
{{7m}\over b^3}({\bf n}\cdot{\bf b})^2-2({\bf n}\cdot{\bf v})^2-
{m\over b}\biggr] \nonumber \\
&{}&
+{2\over r}({\bf n}\cdot{\bf b})\biggl[{{3m}\over b^3}({\bf n}\cdot{\bf b})^2
-6({\bf n}\cdot{\bf v})^2-{m\over b}\biggr] \nonumber \\
&{}&
+{3\over r^2}({\bf n}\cdot{\bf v})\lbrack b^2-5({\bf n}\cdot{\bf b})^2\rbrack
+{1\over r^3}({\bf n}\cdot{\bf b})\lbrack 3b^2-5({\bf n}\cdot{\bf b})^2\rbrack
\biggr\}\,, \nonumber \\
g_{0i}&=&-{{4\mu}\over r} \biggl\{ \biggl[ ({\bf n}\cdot{\bf v})+
{1\over r} ({\bf n}\cdot{\bf b})\biggr]v^i -
{m\over b^3}({\bf n}\cdot{\bf b})b^i\biggr\} \nonumber \\
&{}& +{{2\mu}\over r} {{\delta m}\over m}\biggl( \biggl\{
2({\bf n}\cdot{\bf v})^2-{{3m}\over b^3}({\bf n}\cdot{\bf b})^2+
{6\over r}({\bf n}\cdot{\bf b})({\bf n}\cdot{\bf v}) \nonumber \\
&{}& +{1\over r^2}\lbrack 3({\bf n}\cdot{\bf b})^2-
b^2\rbrack \biggr\}v^i \nonumber \\
&{}& +\biggl\{ -{{4m}\over b^3}({\bf n}\cdot{\bf b})({\bf n}\cdot{\bf v})+
{m\over {rb}}\biggl[1-{3\over b^2}({\bf n}\cdot{\bf b})^2\biggr] \biggr\}b^i
\biggl)\,,\nonumber \\
g_{ij}&=&\delta_{ij}\biggl(1+{{2\tilde{m}} \over r}+{m^2\over r^2} \nonumber \\
&{}&+{\mu \over r} \biggl\{ 2({\bf n}\cdot{\bf v})^2-{{2m}\over b^3}
({\bf n}\cdot{\bf b})^2+{6\over r}({\bf n}\cdot{\bf b})({\bf n}\cdot{\bf v})
+{1\over r^2}\lbrack 3({\bf n}\cdot{\bf b})^2-b^2\rbrack\biggr\}\nonumber \\
&{}&+{\mu \over r} {{\delta m}\over m}\biggl\{ ({\bf n}\cdot{\bf v}) \biggl[
{{7m}\over b^3}({\bf n}\cdot{\bf b})^2-2({\bf n}\cdot{\bf v})^2+
{m\over b}\biggr] \nonumber \\
&{}& +{6\over r}({\bf n}\cdot{\bf b})\biggl[{m\over b^3}
({\bf n}\cdot{\bf b})^2-2({\bf n}\cdot{\bf v})^2\biggr] \nonumber \\
&{}&
+{3\over r^2}({\bf n}\cdot{\bf v})\lbrack b^2-5({\bf n}\cdot{\bf b})^2\rbrack
+{1\over r^3}({\bf n}\cdot{\bf b})\lbrack 3b^2-5({\bf n}\cdot{\bf b})^2
\rbrack \biggl\} \biggl)\nonumber \\
&{}&+{m^2 \over r^2}n^i n^j+{{4\mu}\over r}\biggl(v^i v^j-
{m\over b^3}b^i b^j\biggl)+{{2\mu}\over r} {{\delta m}\over m}\biggl\{
{{6m}\over b^3}({\bf n}\cdot{\bf b})v^{(i} b^{j)}\nonumber \\
&{}& +\biggl[({\bf n}\cdot{\bf v})+{1\over r}({\bf n}\cdot{\bf b})\biggr]
\biggl({m\over b^3} b^i b^j-2 v^i v^j \biggr)\biggr\}\,.
\label{21}
\end{eqnarray}

Our aim is to evaluate definitions (14) and (12) for the gravitational 
energy-momentum and the corresponding fluxes. These definitions are
invariant under global SO(3,1) transformations. Therefore they are frame
dependent. However, $P^a$ is a vector under global Lorentz transformations.
The frame dependence of the gravitational energy-momentum is
understood by simply considering a black hole of mass $m$ and an observer that
is very distant from the black hole. The black hole will appear to this 
observer as a particle of mass $m$, with energy $E=cP^{(0)} =mc^2$ ($m$ is the
mass of the black hole in the frame where the 
black hole is at rest). If, however, the black hole is moving at velocity $v$ 
with respect to the observer, then its total gravitational energy will be 
$E=\gamma m c^2$, where $\gamma=(1-v^2/c^2)^{-1/2}$. This example is a 
consequence of the special theory of relativity, and shows clearly the frame 
dependence of the gravitational energy-momentum. The frame dependence 
is not restricted to observers at spacelike infinity. It holds for observers 
located everywhere in the three-dimensional space.

In order to evaluate definitons (11-14) out of the metric tensor given by (21)
we choose a configuration of tetrad fields that has a clear physical 
interpretation.
In the framework of the TEGR the tetrad field describes both the gravitational
field and the frame. For a given metric tensor there exists an infinity of 
possible frames, and each frame is characterized by six conditions on the 
tetrad field. Three conditions fix the kinematical state of the observer in the
three-dimensional space (for instance, the observer may be stationary in 
space), and the other three conditions fix the orientation of the frame 
(alternatively, the frame may be characterized by the six components of the 
acceleration tensor $\phi_{ab}$ \cite{Maluf5}). 

Therefore tetrad fields are interpreted as reference frames adapted 
to preferred fields of observers in spacetime. This interpretation is possible
by identifying the $e_{(0)}\,^\mu$ components of the frame with the 
four-velocities $u^\mu$ of the observers, $e_{(0)}\,^\mu=u^\mu$ \cite{Maluf5}.
Here we will establish a set of tetrad fields adapted to static observers in 
spacetime. Thus we require $e_{(0)}\,^i=0$. This condition fixes 3 components 
of the frame. The other three components are fixed by choosing an orientation 
of the frame in the three-dimensional space. Therefore $e_{(0)}\,^\mu$ is 
parallel to the worldline of the observers, and  $e_{(k)}\,^\mu$ are the three
unit vectors orthogonal to the timelike direction. We fix $e_{(k)}\,^\mu$ such 
that $e_{(1)}\,^\mu$, $e_{(2)}\,^\mu$ and $e_{(3)}\,^\mu$ in cartesian 
coordinates (and in the flat space-time limit) are unit vectors along the $x$, 
$y$ and $z$ directions. The tetrad field that satisfies these conditions is 
given by

\begin{equation}
e^a\,_\mu(t,x,y,z)=\pmatrix{
A&B&C&0\cr
0&D&E&F\cr
0&0&G&H\cr
0&0&0&I}\,,
\label{22}
\end{equation}
with the following definitions:

\begin{eqnarray}
A&=&(-g_{00})^{1/2}\,, \nonumber \\
B&=&- {{g_{01}} \over {(-g_{00})^{1/2}}}\,, \nonumber \\
C&=&- {{g_{02}} \over {(-g_{00})^{1/2}}}\,,\nonumber \\
D&=&{{\lambda_{11}}\over {(-g_{00})^{1/2}}}\,, \nonumber \\
E&=&{1\over {(-g_{00})^{1/2}}} 
{{\lambda_{12}^2} \over {\lambda_{11}}}\,, \nonumber \\
F&=& (-g_{00})^{1/2}\, {{g_{13}}\over {\lambda_{11}}}\,, \nonumber \\
G&=& {1\over {(-g_{00})^{1/2}}} \biggl[\lambda_{22}^2-
{{\lambda_{12}^4}\over {\lambda_{11}^2}}\biggr]^{1/2}\,, \nonumber \\
H&=& {{(-g_{00})^{1/2}}\over \lambda_{11}}
{{g_{23}\lambda_{11}^2-g_{13}\lambda_{12}^2} \over
{(\lambda_{11}^2 \lambda_{22}^2-\lambda_{12}^4)^{1/2}}}\,,\nonumber \\
I&=&{1\over \lambda_{11}}\biggl[
g_{33}\lambda_{11}^2- (-g_{00})\biggl( g_{13}^2 +
{{(g_{23}\lambda_{11}^2-g_{13}\lambda_{12}^2)^2} \over
{(\lambda_{11}^2\lambda_{22}^2-\lambda_{12}^4)}}\biggr) \biggr]^{1/2}\,.
\label{23}
\end{eqnarray}
The quantity $\lambda_{ij}$ is defined by 
$\lambda_{ij}^2=g_{0i}g_{0j}-g_{00}g_{ij}$, and all metric components are
obtained from (21). In the limit $r\rightarrow \infty$ the asymptotic 
quantities $h_{00}, h_{11}, h_{22}$ and $h_{33}$ are defined by the 
expressions

\begin{eqnarray}
g_{00}&=&-1+h_{00}\,, \nonumber \\
g_{11}&=&1+h_{11}\,, \nonumber \\
g_{22}&=&1+h_{22}\,, \nonumber \\
g_{33}&=&1+h_{33}\,. 
\label{24}
\end{eqnarray}
In terms of these quantities the asymptotic form of the tetrad field is reduced
to 

\begin{equation}
e^a\,_\mu(t,x,y,z)\cong\pmatrix{
1-{{h_{00}}\over 2}&-g_{01}&-g_{02}&0\cr
0&1+{{h_{11}}\over 2}&g_{12}&g_{13}\cr
0&0&1+{{h_{22}}\over 2}&g_{23}\cr
0&0&0&1+{{h_{33}}\over 2}}\,.
\label{25}
\end{equation}
Expression (22) represents a frame that is adapted to static observers 
everywhere in space-time.

\section{Gravitational energy of binary black holes in circular motion}

For a given space-time metric tensor and a given frame, the energy-momentum 
of the space-time is evaluated out of eq. (14). It reads

\begin{equation}
P^a= 4k\oint_S dS_j\,e \Sigma^{a0j} \,.
\label{26}
\end{equation}
If the surface of integration $S$ is fixed at spatial infinity, i.e., 
$S\rightarrow \infty$, $P^a$ yields the total energy-momentum of the 
space-time. The latter is the same for all tetrad fields that exhibit the same
asymptotic behaviour. In particular, the energy-momentum obtained out of frames
that are adapted to static observers at spacelike infinity coincides with the
one obtained out of (22). 

Considering the tetrad field given by eq. (22), the gravitational energy of the
space-time determined by eq. (21) is given by

\begin{eqnarray}
P^{(0)}&=&4ke\oint_S dS_j\,e(e^{(0)}\,_0\Sigma^{00j}+e^{(0)}\,_1\Sigma^{10j}+
e^{(0)}\,_2\Sigma^{20j}) \nonumber \\
&=&4k\oint_SdS_j\, e(A\,\Sigma^{00j}+B\,\Sigma^{10j}+C\,\Sigma^{20j})\,,
\label{27}
\end{eqnarray}
where $A$, $B$ and $C$ are defined by (23), $\Sigma^{\mu\nu\lambda}$ is 
calculated out of (5), and $j=1,2,3$. We will evaluate the expression of 
$P^{(0)}$ for a closed surface $S$ in the asymptotic region $r>>m$ and $r>>b$, 
which characterizes the radiation zone, and then we make $S\rightarrow \infty$.
The metric tensor may be 
decomposed as $g_{\mu\nu}\cong\eta_{\mu\nu}+h_{\mu\nu}$, and $h_{\mu\nu}$ is 
of order $1/r$ at spacelike infinity. Thus we also have 
$g^{\mu\nu}\cong \eta^{\mu\nu}-h^{\mu\nu}$, where 
$h^{\mu\nu}=\eta^{\mu\alpha}\eta^{\nu\beta} h_{\alpha\beta}$. In view of the 
intricate structure of the metric tensor given by (21), in the evaluation of 
(27) we will make the approximation 
$\Sigma^{\mu\nu\lambda}\cong\eta^{\mu\alpha}\eta^{\nu\beta}\eta^{\lambda\gamma}
\Sigma_{\alpha\beta\gamma}$, making sure that $e \Sigma^{a0j}$ is of order 
(at least) $1/r^2$. After a large number of calculations,
and taking into account the approximations explained above, we obtain the exact
expression for the total energy of the space-time,

\begin{eqnarray}
P^{(0)}&=&(m_1+m_2)\biggl[ 1- {{m_1m_2}\over {2(m_1+m_2)b}}\biggr] \nonumber \\
&-&\Big\{\frac{8}{3}\frac{\mu^2m^2}{b^2}\omega\cos(2\omega
\tau)+\frac{\mu^2(\delta m)^2}{b^2}
\left(\frac{m\omega}{b}\right)\Big[\frac{3}{2}\cos(2\omega\tau) \nonumber\\
&+&\frac{3}{8}\cos^4\omega\tau+
\frac{29}{8}\sin^4\omega\tau\Big]\Big\}\sin\omega\tau\cos\omega\tau
\,.
\label{28}
\end{eqnarray}
Assuming that the energy-momentum of matter fields vanishes for the binary
black holes, then the expression above does represent the 
gravitational energy of the space-time. 

We will be interested in average values of time dependent 
quantities. Therefore we define

\begin{equation}
<P^{(0)}>={1\over T}\int_0^T d\tau\, P^{(0)}\,,
\label{29}
\end{equation}
where $T=2\pi/\omega$. We easily obtain

\begin{equation}
<P^{(0)}>= (m_1+m_2)\biggl[ 1- {{m_1m_2}\over {2(m_1+m_2)b}}\biggr]\,.
\label{30}
\end{equation}
We define the binding energy of the configuration according to 
$E_b=<P^{(0)}>-(m_1+m_2)$. We find

\begin{equation}
E_b=-{{m_1m_2}\over {2b}}\,.
\label{31}
\end{equation}
We note that $E_b$ is the the standard non-spinning part of the expression 
for the binding gravitational energy \cite{Wald,Dain} for two black holes in
circular orbit. This term is precisely
the same as the first term in eq. (191) of ref. \cite{Blanchet3}.

We evaluate now the total flux $\Phi^{(0)}=-\dot{P}^{(0)}$ of gravitational
radiation. After a large number of calculations  we arrive at

\begin{eqnarray}
\Phi^{(0)}&=&\Big\{\frac{8}{3}\frac{\mu^2m^2}{b^2}\omega^2\cos(2\omega
\tau)+\frac{\mu^2(\delta
m)^2}{b^2}\left(\frac{m\omega^2}{b}\right)\Big[\frac{3}{2}\cos(2\omega\tau)
+\frac{3}{8}\cos^4\omega\tau \nonumber \\
&+&\frac{29}{8}\sin^4\omega\tau\Big]\Big\}\cos(2\omega\tau)\nonumber\\
&+&\Big\{-\frac{8}{3}\frac{\mu^2m^2}{b^2}(2\omega^2)\sin(2\omega
\tau)+ \frac{\mu^2(\delta
m)^2}{b^2}\left(\frac{m\omega^2}{b}\right)\Big[-3\sin(2\omega\tau)\nonumber \\
&-&\frac{3}{2}\cos^3\omega\tau\sin\omega\tau+ 
\frac{29}{2}\sin^3\omega\tau\cos\omega\tau\Big]\Big\}
\sin\omega\tau\cos\omega\tau\,.
\label{32}
\end{eqnarray}
An interesting consequence of the expression above is that the average value 
of $\Phi^{(0)}$ over a complete cycle vanishes,

\begin{equation}
<\Phi^{(0)}>=0\,.
\label{33}
\end{equation}
We conclude that the orbital (stationary) motion of two black holes (or two 
point masses) on a plane produce the gravitational radiation given by eq. 
(32). However, the average value of this radiation vanishes.
On the other hand, the 
situation changes if the separation distance $b$ changes with time.

In order to obtain (32) we have considered that $b$ is not a function of $t$.
This condition was assumed in ref. \cite{Alvi}. According to ref. \cite{Owen},
however, this condition may be relaxed. Thus we may admit that 
$b$ varies slowly with time. Taking into account this 
assumption we obtain an expression for the total flux of the gravitational 
radiation. For arbitrary $b(t)$ we have $\dot \omega \ne 0$. Consequently 
neither a complete cycle nor a period can be defined. In fact, the concept 
of average value does not apply to this case. 
$\Phi^{(0)}$ is now clearly nonvanishing.  We present here only the 
contribution to $\Phi^{(0)}$ of the variation in time of the first term on the 
right hand side of eq. (28), namely, the variation in time of the binding 
energy, which does not even contribute to (32). We find

\begin{equation}
\Phi^{(0)}\approx-{{d E_b} \over { dt}}  =-{{m_1m_2}\over {2b^2}}\dot{b}\,.
\label{34}
\end{equation}
This flux will be positive definite provided $\dot{b}<0$, which is the case
for realistic binary black holes prior to the merger.

We have evaluated the 
momentum component $P^{(3)}$, which is the component oriented along the $z$ 
direction. The integration is carried out over a finite cubic volume with
sides $a$, such that $a>> b$. We found that $P^{(3)}=0$. This result is
expected since the black holes are restricted to the $xy$ plane, and 
there is no flux of momentum along the $z$ direction, in contrast to a general
situation to be addressed in the next section. When the integration is
carried out over the whole three-dimensional space, the total momentum of the
space-time vanishes, i.e., $P^{(i)}=0$ for $i=1,2,3$.

\section{Arbitrary time dependent metric tensor}

The standard form of the metric tensor for the space-time of the inspiral of
two black holes is given by \cite{Tagoshi,Faye,Thorne} 

\begin{eqnarray}
g_{00}&=& -1+2V-2V^2+8X \nonumber \\
g_{0i}&=& -4V_i-8R_i \nonumber \\
g_{ij}&=&\delta_{ij}(1+2V+2V^2)+4W_{ij}\,.
\label{35}
\end{eqnarray}
The form of the potentials $V, V_i, X, R_i$ and $W_{ij}$ may be obtained in the 
three references indicated above. Here we will just assume that $V$, $V_i$, 
$R_i$ and $X$ behave as $1/r$ at spacelike infinity, whereas $W_{ij}$ behaves
as $1/r^2$. The expression of the energy-momentum $P^a$ will be given in terms 
of these potentials. The use of the explicit form of the potentials yields a 
rather intricate form of the energy-momentum. The spinning nature of the 
solution is manifest in the potentials $V, V_i$ and $W_{ij}$ \cite{Faye}.

In cartesian coordinates a stationary observer in space-time is described
by a frame very similar to eq. (22). In the metric tensor given by (21) we have
$g_{03}=0$, which is not the case for (35). The frame that (i) yields (35), 
(ii) is
adapted to stationary observers in space-time, i.e., $e_{(0)}\,^i=0$, and (iii)
is oriented along the $x,y,z$ directions at spacelike infinity, i.e., 
$e_{(i)}\,^j(t,x,y,z)\cong \delta^j_i$ when $r\rightarrow \infty$,  is given by

\begin{equation}
e^a\,_\mu= \left(
             \begin{array}{cccc}
             A & B & C & J \\
             0 & D & E & F \\
             0 & 0 & G & H \\
             0 & 0 & 0 & I \\
             \end{array}
             \right)\,,
\label{36}
\end{equation}
where the following relations are satisfied,

\begin{eqnarray}
A^2&=&-g_{00} \nonumber \\
AB&=& -g_{01} \nonumber \\ 
AC&=& -g_{02} \nonumber \\
AJ&=& -g_{03} \nonumber \\
-B^2+D^2&=& g_{11} \nonumber \\
-BC+DE&=& g_{12} \nonumber \\
-BJ+FD&=& g_{13} \nonumber \\
-C^2+E^2+G^2&=& g_{22} \nonumber \\
-CJ+EF+GH&=& g_{23} \nonumber \\
-J^2+F^2+H^2+I^2&=&g_{33}\,. 
\label{37}
\end{eqnarray}
The relations above allow to obtain all tetrad components in terms of the
metric tensor components.

We present below all components of the energy-momentum obtained out of the
tetrad field (36), assuming the asymptotic behaviour of the potentials as
explained above. We have discarded several terms that 
fall off as $O(1/r^3)$ or faster. For $P^{(0)}$ we obtain

\begin{eqnarray}
P^{(0)}& =&-k\Big\{\lim_{x\rightarrow\pm\infty}\int_{-\infty}^{\infty}dydz
\left[\partial_1(g_{22}+g_{33})\right] \nonumber \\
&+&\lim_{y\rightarrow\pm\infty}\int_{-\infty}^{\infty}dxdz
\left[\partial_2(g_{11}+g_{33})\right]\nonumber\\
&+&\lim_{z\rightarrow\pm\infty}\int_{-\infty}^{\infty}dxdy
\left[\partial_3(g_{11}+g_{22})\right]\nonumber \\
&+&\lim_{x\rightarrow\pm\infty} \int_{-\infty}^{\infty}dydz
\left[C\partial_0(g_{12})+J\partial_0(g_{13})\right]\nonumber\\
&+&\lim_{y\rightarrow\pm\infty}\int_{-\infty}^{\infty}dxdz
\left[B\partial_0(g_{12})+J\partial_0(g_{23})\right] \nonumber \\
&+&\lim_{z\rightarrow\pm\infty}\int_{-\infty}^{\infty}dxdy
[B\partial_0(g_{13})+C\partial_0(g_{23})]\nonumber\\
&+&\lim_{x\rightarrow\pm\infty}\int_{-\infty}^{\infty}dydz B\partial_0
(g_{11})+\lim_{y\rightarrow\pm\infty}\int_{-\infty}^{\infty}dxdz C\partial_0
(g_{22})\nonumber \\
&+&\lim_{z\rightarrow\pm\infty}\int_{-\infty}^{\infty}
dxdy J\partial_0 (g_{33})\Big\}\,,
\label{38}
\end{eqnarray}
where

{\small
$$\lim_{x\rightarrow\pm\infty}\int_{-\infty}^{\infty}
dydz F(x,y,z)= \int_{-\infty}^{\infty} dy 
\int_{-\infty}^{\infty} dz\, F(\infty, y,z)- 
\int_{-\infty}^{\infty} dy 
\int_{-\infty}^{\infty}dz F(-\infty, y,z)\,.$$}
Equation (38) reproduces (28) if we reduce the metric tensor (35) to the
form given by (21), in which case we make $J=0$.

In view of the asymptotic behaviour of 
the metric tensor (35), the total momentum of the space-time vanishes. 
In the expressions below for $P^{(i)}$, we formally integrate over 
a finite surface $S_0$ of a large rectangular
volume with sides $(2x_0,2y_0,2z_0)$, such that $x_0=y_0=z_0=a$, and
$a$ is much larger than the separation of the black holes. We arrive at

\begin{eqnarray}
P^{(1)}&=&-32k\Biggl[\int_{S_0}dy\,dz
(V_{1}\partial_0V_{1})+\int_{S_0}
dz\,dx(V_{1}\partial_0V_{2})\nonumber \\
&+&\int_{S_0}dx\,dy (V_{1}\partial_0V_{3})\Biggr]
\nonumber \\
&+&4k\Biggl[\int_{S_0}dy\,dz\,\partial_0(V+W_{22}+W_{33})
\nonumber \\
&-&\int_{S_0} dx\,dy(\partial_0W_{13})
-\int_{S_0}dz\,dx(\partial_0W_{12})\Biggr]\,,
\label{39}
\end{eqnarray}

\begin{eqnarray}
P^{(2)}&=&-32k\Biggl[\int_{S_0}dy\,dz
(V_{2}\partial_0V_{1})+\int_{S_0}
dz\,dx(V_{2}\partial_0V_{2})\nonumber \\
&+&\int_{S_0}dx\,dy (V_{2}\partial_0V_{3})\Biggr]
\nonumber \\
&+&4k\Biggl[\int_{S_0}dz\,dx\,\partial_0(V+W_{11}+W_{33})
\nonumber \\
&-&\int_{S_0}dy\,dz(\partial_0W_{12})-\int_{S_0}
dx\,dy(\partial_0W_{23})\Biggr]\,,
\label{40}
\end{eqnarray}

\begin{eqnarray}
P^{(3)}&=&-32k\Biggl[\int_{S_0}dy\,dz
(V_{3}\partial_0V_{1})+\int_{S_0}
dz\,dx(V_{3}\partial_0V_{2})\nonumber \\
&+&\int_{S_0}dx\,dy (V_{3}\partial_0V_{3})\Biggr]
\nonumber \\
&+&4k\Biggl[\int_{S_0}dy\,dz\,\partial_0(V+W_{11}+W_{22})
\nonumber \\
&-&\int_{S_0}dz\,dx(\partial_0W_{23})-\int_{S_0}
dy\,dz(\partial_0W_{13})\Biggr]\,,
\label{41}
\end{eqnarray}
where

$$\int_{S_0}dy dz\, F(x,y,z)=\int_{-y_0}^{y_0}dy 
\int_{-z_0}^{z_0} dz\, F(x_0,y,z)-\int_{-y_0}^{y_0} dy
\int_{-z_0}^{z_0} dz\,F(-x_0,y,z)\,,$$
etc. An important conclusion that we can draw from the expressions above is
that the potentials $X$ and $R_i$ do not contribute to the momenta.

Considering the post-Newtonian potentials $V, V_i$ and $W_{ij}$, we 
find that the $P^{(3)}$ component of the gravitational momentum is, in
general, nonvanishing for a finite volume of the three-dimensional space,
and exhibits a dependence in time. Therefore, it yields a momentum 
flux $\phi^{(3)}=-\dot{P}^{(3)}$ which, in turn, is likely to be related to
the bobbing of the black holes. Unfortunately the post-Newtonian potentials
of ref. \cite{Tagoshi} are not suitable for the present analysis, because they
are valid only in the near zone, and we are interested in the expressions of 
the potentials in the radiation zone. Some of the potentials presented in the 
latter reference diverge with the increasing of the radial distance $r$, a 
feature that prevents us from calculating all momentum components. 

The post-Newtonian potentials $V, V_i$ and $W_{ij}$ depend on the spins 
$S_x$, $S_y$ and $S_z$ of the black holes, which are time dependent functions
\cite{Faye}. The explicit form of these functions, for given initial 
conditions, and an analytic, widely accepted 
form of the post-Newtonian potentials, are not 
available in the literature. For this reason, we cannot proceed and obtain the
detailed form of expressions (39), (40) and (41).

\section{Concluding remarks}

In this paper we have analyzed the metric tensor for the nonspinning black hole
binary in circular orbit in the $xy$ plane, in the context of the teleparallel 
equivalent of general relativity. This metric tensor is an approximate solution 
of Einstein's equations in which the distance between the holes is constant in 
time. 

We have also addressed the general post-Newtonian form of the metric tensor 
that describes the inspiral and merger of two spinning black holes. The
total energy-momentum of the space-time may be expressed in a simple form in
terms of the metric tensor components, and may be easily computed provided 
the post-Newtonian potentials are given. We have found that only the potentials
$V$, $V_i$ and $W_{ij}$ contribute to the momenta. 
It is very likely that the time dependence of the momentum component $P^{(3)}$ 
(the momentum component oriented along the $z$ direction) is related to the
bobbing of the spinning black holes with oppositely directed spins restricted
to the orbital plane. 

The calculations of the total gravitational energy, and the corresponding flux
for the nonspinning black hole binary in circular orbit, yield a quite 
interesting result. The total energy $P^{(0)}$ and the gravitational flux 
$\phi^{(0)}$ are given by eqs. (28) and (32). The former yields the known 
result for the mean value of the binding energy of the configuration, whereas
the average value of $\phi^{(0)}$ in time vanishes, $<\phi^{(0)}>=0$.
It means that for two black holes in circular 
orbit, the average value of the gravitational radiation is zero. This 
result is consistent with the stationary character of the space-time. A
nontrivial emission of gravitational radiation must necessarily be related to
a loss of energy-momentum of the source (as the loss of mass described by
Bondi's radiating metric \cite{Bondi}). If, however, the separation between 
the holes decreases in time, as in an actual evolution of the black hole
binary, a nonvanishing (definite positive) flux of gravitational radiation is 
emitted. 

This result is conceptually different from the conclusion drawn from 
Eddington's spinning rod \cite{Eddington}, which was reconsidered by other
authors \cite{Cooperstock}. In the framework of pseudotensor definitions and 
of linearized general relativity, the quadrupole formula was 
obtained. The latter relates the energy loss of the system with the 
variation in time of the mass-quadrupole 
of the source. This formula was used by Eddington to deduce 
the energy flux generated by a rod that spins in the $xy$ plane with 
angular frequency $\omega$. Let $I$ represent the moment of inertia of the
rod, and $G$ the gravitational constant. The total energy flux is given by
\cite{Eddington,Landau}

\begin{equation}
{{dE}\over {dt}}=-{{32 G I^2 \omega^6}\over
{5c^5}}\,.
\label{42}
\end{equation}
In contrast to the approach that allows to deduce the formula above, we 
note that the procedure that led to eqs. (32) and (33) is based neither on 
pseudotensors nor on the linearized form of the theory. Equation (10) is a
true tensorial quantity. It is valid for any coordinate system of the 
three-dimensional spacelike hypersurface, and for finite volume $V$ and 
corresponding surface $S$. We 
believe that the reconsideration of Eddington's spinning rod in the present
context would also lead to an equation of the type $<\phi^{(0)}>=0$,
provided the angular frequency is constant.

\end{document}